\documentclass[11pt]{birk2}
\def\blackbox{\hfill\rule{.1in}{.1in}}
\def\sumin{\sum_{i=1}^n}
\def\tr{{\rm t}}
\def\E{{\rm E}}
\def\Var{{\rm Var}}
\def\var{{\rm var}}
\def\d{{\rm d}}
\def\N{{\rm N}}
\def\full{{\rm full}}

\def\sqb{{\rm sqb}}
\def\mse{{\rm mse}}
\def\Tr{{\rm Tr}}
\def\midd{\,|\,}

\def\hatt{\widehat}
\def\tilda{\widetilde}
\def\rootn{\sqrt{n}}
\def\quadandquad{\quad{\rm and}\quad}
\def\arr{\rightarrow}

\def\Gam{{\rm Gamma}}
\def\fic{{\rm FIC}}
\def\wfic{{\rm wFIC}}
\def\EE{{\cal E}}
\def\whatsqb{\hbox{\rm w-$\hatt\sqb$}}

\def\whatvar{\hbox{\rm w-$\hatt\var$}}

\def\hop{\medskip}

\def\beq{\begin{eqnarray}}
\def\eeq{\end{eqnarray}}
\def\beqn{\begin{eqnarray*}}
\def\eeqn{\end{eqnarray*}}

\begin{document}

\author{Nils Lid Hjort \\
\medskip
{\it Department of Mathematics, University of Oslo}}


\chapter{Focussed Information Criteria \\ 
for the Linear Hazard Regression Model}

\markboth{Nils Lid Hjort}{FIC for the linear hazard regression model}

\noindent{\bf Abstract:} 
The linear hazard regression model developed by Aalen
is becoming an increasingly popular alternative
to the Cox multiplicative hazard regression model. 
There are no methods in the literature for 
selecting among different candidate models of 
this nonparametric type, however. In the present paper 
a focussed information criterion is developed for 
this task. The criterion works for each specified 
covariate vector, by estimating the mean squared
error for each candidate model's estimate of
the associated cumulative hazard rate; the finally
selected model is the one with lowest estimated
mean squared error. Averaged versions of the criterion
are also developed. 

\bigskip

\noindent{\bf Keywords and phrases:} 
Aalen's linear model, 
covariate selection,
focussed information criterion,
hazard regression,
model selection 


\vspace*{30pt}
\hrule
\section{Introduction: Which Covariates to Include?}
\label{section:intro}

We consider survival regression data of the usual form
$(T_i,\delta_i,x_i)$ for individuals $i=1,\ldots,n$,
where $x_i$ is a vector of say $r$ covariates, 
among which one wishes to select those of highest
relevance. Also, $T_i=\min\{T_i^0,C_i\}$ is the
possibly censored life-length and $\delta_i=I\{T_i^0<C_i\}$
the associated non-censoring indicator, in terms
of underlying life-length $T_i^0$ and censoring time
$C_i$ for individual $i$. 

Our framework is that of the linear hazard regression
model introduced by Aalen (1980),
see e.g.~the extensive discussion in 
Andersen, Borgan, Gill and Keiding (1993, Ch.~8) 
and Martinussen and Scheike (2006, Ch.~5), 
where the hazard rate for individual $i$ may be represented as 
$$h_i(u)=x_i^\tr\alpha(u)
   =\sum_{j=1}^r x_{i,j}\alpha_j(u) 
  \quad {\rm for\ }i=1,\ldots,n, $$
in terms of regressor functions $\alpha_1(u),\ldots,\alpha_r(u)$. 
These need to satisfy the requirement that 
the linear combination $x^\tr\alpha(u)$ stays nonnegative
for all $x$ supported by the distribution of covariate
vectors. In other words, the associated cumulative
hazard function 
\beq
\label{eq:Htx}
H(t\midd x)=\int_0^t x^\tr\alpha(u)\,\d u
   =x^\tr A(u)
   =\sum_{j=1}^r x_jA_j(t) 
\eeq
is nondecreasing in $t$, for all $x$ in the relevant
covariate space; here we write 
$A_j(t)=\int_0^t\alpha_j(u)\,\d u$ for $j=1,\ldots,r$. 

Among questions discussed in this paper is when we might
do better with only a subset of the $x$ covariates 
than with keeping them all. We focus specifically
on the problem of estimating $H(t\midd x)$ of 
(\ref{eq:Htx}) well, for a specified individual 
carrying his given covariate information $x$.
The full-model estimator
\beq
\label{eq:Hhatfull}
\hatt H(t\midd x)=\hatt H_\full(t\midd x)=x^\tr\hatt A(t)
   =\sum_{j=1}^r x_j\hatt A_j(t) 
\eeq
is one option, using the familiar Aalen estimators
for $A_1,\ldots,A_r$ in the full model, 
keeping all covariates on board. 
Pushing some covariates out of the model leads to competing
estimators of the type
\beq
\label{eq:HtildeI}
\tilda H_I(t\midd x)=\sum_{j\in I}x_j\tilda A_{I,j}(t), 
\eeq
where the index set $I$ is a subset of $\{1,\ldots,r\}$,
representing those covariates that are kept 
in the model, and where the $\tilda A_{I,j}(t)$'s
for $j\in I$ are the Aalen estimators in 
the linear hazard rate model associated with the 
$I$ covariates. Using $\tilda H_I(t\midd x)$ 
instead of $\hatt H(t\midd x)$ will typically correspond
to smaller variances but to modelling bias.
Slightly more generally, bigger index sets $I$
imply more variance but less modelling bias,
and vice versa. Thus the task of selecting 
suitable covariates amounts to a statistical
balancing game between sampling variability and bias. 

In Section \ref{section:submodels} we fix the framework
and give proper definitions of full-model 
and submodel estimators. These are also expressed 
in terms of counting processes and at-risk processes.
Links with martingale theory make it possible
in Section \ref{section:biasvariance} to 
accurately assess the bias and variance properties
associated with a given candidate model. 
This is followed up in Section \ref{section:estimatingrisks} 
by explicit methods for estimating bias and variance
from the data. The focussed information criterion
(FIC) introduced in Section \ref{section:fic} 
acts by estimating the risk associated with 
each candidate model's estimator of the cumulative
hazard function; the model we suggest being used 
in the end is the one with lowest estimated risk.
Weighted versions are also put forward. 
In an extended version of the present work 
I shall report on the use of the methods 
for real data and in some simulation setups. 
The present paper ends with a list of concluding remarks
in Section \ref{section:concludingremarks}. 

The brief introduction has so far taken model comparison
as corresponding to accuracy of estimators of cumulative
hazard rates $H(t\midd x)$. By a delta method argument
this is also nearly equivalent to ranking models in terms of
accuracy of estimates of survival probabilities
$S(t\midd x)=\exp\{-H(t\midd x)\}$, 
where the estimates in question take the form
$$\hatt S_\full(t\midd x)=\prod_{[0,t]}\{1-x^\tr\,\hatt A(u)\}
  \quadandquad
  \tilda S_I(t\midd x)=\prod_{[0,t]}
   \Bigl\{1-\sum_{j\in I}x_j\,\d\tilda A_{I,j}(u)\Bigr\}. $$
(For details regarding notation for and properties of 
the product integral used on the right, see 
e.g.~Andersen et al.~(1993, Ch.~II.6).)
It is important to realise that a submodel $I$ may
work better than the full model, even if the submodel
in question is not `fully correct' as such;
this is determined, among other aspects, 
by the sizes of the $\alpha_j(u)$ regressor functions
that are left out a model. 
This makes model selection different in spirit
and operation than e.g.~performing goodness-of-fit 
checks on all candidate models. 

Aalen's linear hazard model is in many
important respects different from Cox's proportional
hazard model, also regarding the mathematical 
treatment of estimators and their properties; 
see Andersen et al.~(op.~cit.). 
We note that focussed information criteria 
and a general theory for model averaging estimators 
for the Cox model have been developed 
in Hjort and Claeskens (2006). Based on research 
in that and in the present paper methods 
may be devised that can help select among 
`the best Cox model' and `the best Aalen model',
in situations where that question is of relevance,
but that theme is not pursued here. 

\vspace*{30pt}
\hrule
\section{Estimators in Submodels}
\label{section:submodels}

This section properly defines the Aalen estimators
$\hatt A$ and $\tilda A_I$ involved in 
(\ref{eq:Hhatfull}) and (\ref{eq:HtildeI}).
It is convenient to define these in terms of 
the counting process and at-risk process
$$N_i(t)=I\{T_i\le t,\delta_i=1\}
  \quadandquad
  Y_i(u)=I\{T_i\ge u\} $$
for individuals $i=1,\ldots,n$. We shall also need
the martingales 
$M_i(t)=N_i(t)-\int_0^t Y_i(u)\,\d H_i(u)$, for which
\beq
\label{eq:dNi}
\d N_i(u)=Y_i(u)x_i^\tr\,\d A(u)+\d M_i(u). 
\eeq
These are orthogonal and square integrable with 
variance processes
\beq
\label{eq:MiMi}
\langle M_i,M_i\rangle(t)
   =\int_0^t Y_i(u)h_i(u)\,\d u
   =\int_0^t Y_i(u)x_i^\tr\,\d A(u). 
\eeq
In other words, $M_i(t)^2-\langle M_i,M_i\rangle(t)$
is another zero-mean martingale, implying
in particular that the mean of (\ref{eq:MiMi})
is equal to the variance of $M_i(t)$. 

Now introduce the $r\times r$-size matrix function
\beq
\label{eq:Gn}
G_n(u)=n^{-1}\sumin Y_i(u)x_ix_i^\tr. 
\eeq
The Aalen estimator $\hatt A=(\hatt A_1,\ldots,\hatt A_r)^\tr$
in the full model corresponds to 
$$ \d\hatt A(u)=G_n(u)^{-1}n^{-1}\sumin x_i\,\d N_i(u), $$
with integrated version 
\beq
\label{eq:aalen}
\hatt A(t)=\int_0^t G_n(u)^{-1}n^{-1}\sumin x_i\,\d N_i(u)
  \quad {\rm for\ }t\ge0. 
\eeq
This also defines $\hatt H_\full(t\midd x)$
of (\ref{eq:Hhatfull}). It is assumed here 
that at least $r$ linearly independent covariate
vectors $x_i$ remain in the risk set at time $t$, 
making the inverse of $G_n$ well-defined for 
all $u\le t$; this event has probability growing 
exponentially quickly to 1 as sample size
increases, under mild conditions. 

To properly define the competitor $\tilda H_I(t\midd x)$ 
of (\ref{eq:HtildeI}), we use the notation 
$x_I=\pi_Ix$ for the vector of those $x_j$ components 
for which $j\in I$, for each given subset $I$
of $\{1,\ldots,r\}$. In other words, $\pi_I$ is the 
projection matrix of size $|I|\times r$, with 
$|I|$ the number of covariates included in $I$. 
For the given $I$, we partition the $G_n$ function
into blocks, 
$$G_n(u)=\pmatrix{G_{n,00}(u), &G_{n,01}(u) \cr
                  G_{n,10}(u), &G_{n,11}(u) \cr}, $$
where 
$$G_{n,00}(u)=\pi_IG_n(u)\pi_I^\tr
   =n^{-1}\sumin Y_i(u)x_{i,I}x_{i,I}^\tr $$
is of size $|I|\times|I|$, while $G_{n,11}(u)$ is 
of size $q\times q$ with $q=r-|I|$, etc. 
The Aalen estimator for the vector of $A_j$ functions 
where $j\in I$ is 
$$\tilda A_I(t)=\int_0^t G_{n,00}(u)^{-1}
   n^{-1}\sumin x_{i,I}\,\d N_i(u). $$
These are those at work in (\ref{eq:HtildeI}). 

Using (\ref{eq:dNi}) we may write 
$$n^{-1}\sumin x_{i,I}\,\d N_i(u)
   =n^{-1}\sumin Y_i(u)x_{i,I}x_i^\tr\,\d A(u)
   +n^{-1}\sumin x_{i,I}\,\d M_i(u), $$
which further leads to 
\beq
\label{eq:dtildeAIu}
\begin{array}{rcl}
\d\tilda A_I(u)
&=&\displaystyle 
G_{n,00}(u)^{-1}
   \Bigl\{G_{n,00}(u)\,\d A_I(u)+G_{n,01}(u)\,\d A_{II}(u) \\
& &\displaystyle 
\qquad\qquad
   +\,n^{-1}\sumin x_{i,I}\,\d M_i(u)\Bigr\}, 
\end{array}
\eeq
along with its integrated version. 
Here $II=I^c$ is the set of indexes not in $I$. 
This representation, basically in terms of 
a mean term plus martingale noise,   
is used in the next section to characterise
means and variances of the (\ref{eq:HtildeI}) 
estimators. It again assumes that the $G_n$ 
is invertible on $[0,t]$, an event having 
probability growing exponentially to 1
and therefore not disturbing the main analysis. 

We remark that when the $I$ model is used, 
then the Aalen estimator $\tilda A_I(t)$ does not 
directly estimate $A_I$, but rather the function 
$A_I(t)+\int_0^ t G_{00}^{-1}G_{01}\,\d A_{II}$.  

\vspace*{30pt}
\hrule
\section{Bias, Variance, \\ and Mean Squared Error Calculations}
\label{section:biasvariance}

In this section we develop useful approximations for
the mean squared error of each of the 
(\ref{eq:HtildeI}) estimators 
$\tilda H_I(t\midd x)=x_I^\tr\tilda A_I(t)$.
We shall assume that the censoring variables
$C_1,\ldots,C_n$ are i.i.d.~with some survival
distribution $C(u)=\Pr\{C_i\ge u\}$, and 
that they are independent of the life-times $T_i^0$; 
the case of no censoring corresponds to 
$C(u)=1$ for all $u$. 
It will furthermore be convenient to postulate 
that $x_1,\ldots,x_n$ stem from some distribution 
in the space of covariate vectors. These assumptions
imply for example that the $G_n$ function of 
(\ref{eq:Gn}) converges with increasing sample size, say
\beq
\label{eq:GntoG}
G_n(u)\arr G(u)=\E_* Y(u)xx^\tr
   =\E_*\exp\{-x^\tr A(u)\}xx^\tr\,C(u),
\eeq
where $\E_*$ refers to expectation under the 
postulated covariate distribution. Also the mean function
\beqn
\bar G_n(u)=\E\,G_n(u)
   =n^{-1}\sumin p_i(u)x_ix_i^\tr 
\eeqn
converges to the same limit $G(u)$; here 
$p_i(u)=\E Y_i(u)=\exp\{-x_i^\tr A(u)\}\,C(u)$. 
We shall finally assume that the $r\times r$-function 
$G(u)$ is invertible over the time observation 
window $u\in[0,\tau]$ of interest; this corresponds
to $C(\tau)$ positive and to a non-degenerate 
covariate distribution. 
As in Section \ref{section:submodels} 
there will be a need to partition the $G(u)$
function into blocks $G_{00}(u),G_{01}(u)$, etc.; 
$G_{00}(u)$ has e.g.~size $|I|\times|I|$. 
A similar remark applies to $\bar G_n(u)$. 

Consider as in Section \ref{section:intro} a given
individual with covariate information $x$. 
From representation (\ref{eq:dtildeAIu}), 
\beqn
x_I^\tr\,\d\tilda A_I(u)
&=& x_I^\tr\,\d A_I(u)
   +x_I^\tr G_{n,00}(u)^{-1}G_{n,01}(u)\,\d A_{II}(u) \\
& & \qquad\qquad
   +\,n^{-1/2}x_I^\tr G_{n,00}(u)^{-1}\,\d V_{n,I}(u) \\
&=& x^\tr\,\d A(u)+b_{I,n}(u)^\tr\,\d A_{II}(u)
   +n^{-1/2}x_I^\tr G_{n,00}(u)^{-1}\,\d V_{n,I}(u), 
\eeqn 
in which $V_n$ is the $r$-dimensional martingale process 
with increments
\beq
\label{eq:dVnu}
\d V_n(u)=n^{-1/2}\sumin x_i\,\d M_i(u),  
\eeq
whereas $b_{I,n}$, defined by 
\beq
\label{eq:bnu}
b_{I,n}(u)=G_{n,10}(u)G_{n,00}(u)^{-1}x_I-x_{II}, 
\eeq
can be seen as a bias function 
(omitting at the moment $x$ in the notation for this function).
Its dimension is $q=r-|I|$.  
This leads to the representation 
\beq
\label{eq:basic}
\rootn\{x_I^\tr\tilda A_I(t)-x^\tr A(t)\}
   =\rootn\int_0^t b_{I,n}^\tr\,\d A_{II}
   +x_I^\tr\int_0^t G_{n,00}^{-1}\,\d V_{n,I}.
\eeq
The second term is a zero-mean martingale 
while the first term is a bias term, stemming 
from using model $I$ that does not include all
the components. We shall use (\ref{eq:basic})
to develop good approximations to 
$$\mse_n(I)=\mse_n(I,t)=n\,\E\{\tilda H_I(t\midd x)
   -H(t\midd x)\}^2, $$
the normalised mean squared error of the 
(\ref{eq:HtildeI}) estimator. We treat the covariate
vectors $x_1,\ldots,x_n$ as given, i.e.~our
approximations are expressed directly in terms
of these. 

In view of the assumptions made in the beginning
of this section, a first-order approximation 
to the mean of (\ref{eq:basic}) is 
$\rootn\int_0^t\bar b_{I,n}^\tr\,\d A_{II}$,
since the second term has zero mean; 
here $\bar b_{I,n}(u)=\bar G_{n,10}(u)\bar G_{n,00}(u)^{-1}$. 
Also, $\int_0^t b_{I,n}^\tr\,\d A_{II}$
and $\int_0^t \bar b_{I,n}^\tr\,\d A_{II}$
are both close to the limit $\int_0^t b_I^\tr\,\d A_{II}$,
with high probability for large $n$, where 
$b_I(u)=G_{10}(u)G_{00}^{-1}x_I-x_{II}$. 

To study the second term of (\ref{eq:basic}), note that
$V_n$ of (\ref{eq:dVnu}) is a zero-mean martingale
with variance process
$\langle V_n,V_n\rangle(t)=J_n(t)$, 
with $r\times r$-matrix increments 
$$\d J_n(u)=n^{-1}\sumin Y_i(u)x_ix_i^\tr\,x_i^\tr\,\d A(u). $$
There is a well-defined limit function $J(u)$ with 
increments 
$$\d J(u)=\E_*Y(u)xx^\tr\,x^\tr\,\d A(u)
   =\E_*\exp\{-x^\tr A(u)\}xx^\tr\,x^\tr\,\d A(u)\,C(u) $$
under the conditions stated above. Thus $V_n$ 
converges in distribution to a Gau\ss ian martingale $V$
with increments $\d V(u)$ having zero mean 
and variance matrix $\d J(u)$. It also follows that 
the second term of (\ref{eq:basic}) converges 
in distribution; 
$$x_I^\tr\int_0^t G_{n,00}^{-1}\,\d V_{n,I}
   \arr_d x_I^\tr\int_0^t G_{00}^{-1}\,\d V_I, $$
which is normal with variance 
$$\var(I,t)=x_I^\tr\int_0^t G_{00}^{-1}\,\d J_{00}\,
   G_{00}^{-1}x_I. $$
The integral here is defined in the appropriate 
and natural Riemannian sense, and is also equivalent
to a finite sum of ordinary integrals, 
found by writing out the quadratic form. 

The first term of (\ref{eq:basic}) is essentially
non-random when compared with the second term. 
A more formal statement can be put forward in
a framework of local asymptotic neighbourhoods,
where $\d A_{II}(u)=\d D(u)/\rootn$, say; in this case,
$$\rootn\{\tilda H_I(t\midd x)-H(t\midd x)\}
   \arr_d \int_0^t b(u)^\tr\,\d D(u)
   +\N(0,\var(I,t)). $$
Our main use of these considerations is the 
approximation to the normalised mean squared error; 
\beq
\label{eq:sqbplusvar}
\mse_n(I,t)\doteq\sqb(I,t)+\var(I,t),
\eeq
where $\var(I,t)$ is defined above and 
$$\sqb(I,t)=
   n\Bigl(\int_0^t\bar b_{I,n}^\tr\,\d A_{II}\Bigr)^2. $$

\medskip
{\sc Remark.}
There are often situations where it pays off to
exclude some covariates, even though their 
associated $\alpha_j(u)$ functions are non-zero.
This is a consequence of the squared bias versus variance
balancing game. For example, a submodel $I$ is better
than the full set, for the given covariate $x$, if 
$\sqb(I,t)+\var(I,t)\le0+\var(\full,t)$,
which translates to
$$n\Bigl\{\int_0^t(G_{10}G_{00}^{-1}x_I-x_{II})^\tr\,\d A_{II}\Bigr\}^2
   \le x^\tr\int_0^t G^{-1}\,\d J\,G^{-1}\,x
   -x_I^\tr\int_0^t G_{00}^{-1}\,\d J_{00}\,G_{00}^{-1}\,x_I. $$
This effectively describes a `tolerance radius' around
a given model, inside which the model is preferable
to the full model, even when not perfectly valid. 
The inequality says that a certain linear combination
of the $\alpha_j(u)$ functions for $j\notin I$
should not be too big, compared also to the sample size;
for large $n$ even small biases are costly, and the
full model becomes preferable.~\blackbox

\vspace*{30pt}
\hrule
\section{Estimating the Risks}
\label{section:estimatingrisks}

We have seen that each candidate model $I$ 
has an associated risk $\mse_n(I,t)$ of (\ref{eq:sqbplusvar}) 
when estimating the cumulative hazard function
using $\tilda H_I(t\midd x)$. Here we deal with
the consequent task of estimating these risk
quantities from data. 

For the variance part we use 
$$\hatt\var(I,t)
   =x_I^\tr\int_0^t G_{n,00}^{-1}(u)\,\d\hatt J_{n,00}(u)\,
   G_{n,00}(u)^{-1}\,x_I, $$
wherein 
$$\d\hatt J_n(u)=n^{-1}\sumin Y_i(u)x_ix_i^\tr\,x_i^\tr\d\hatt A(u), $$
engaging the full-model Aalen estimator. 
The $|I|\times|I|$ block used for the variance estimation is 
$\pi_I\,\d\hatt J_n(u)\pi_I^\tr$. 

For the squared bias part, consider in general terms
the quantity $\beta^2$, where $\beta=\int_0^t g^\tr\,\d A_{II}$,
for a specified $q$-dimensional function $g$; 
again, $q=r-|I|$. Considering 
$\hatt\beta=\int_0^t g^\tr\,\d\hatt A_{II}$, 
employing the $II$ part of the full-model Aalen estimator, 
we have 
$$\E\,\hatt\beta\doteq\beta
   \quadandquad
  \Var\,\hatt\beta\doteq n^{-1}\int_0^t g(u)^\tr\,\d Q(u)\,g(u), $$
from results above, where we write 
$$\d Q(u)=\{G(u)^{-1}\,\d J(u)\,G(u)^{-1}\}_{11} $$
for the lower right hand $q\times q$ block of the matrix within brackets,
the block associated with subset $II=I^c$. Thus 
$\E\,\hatt\beta^2\doteq\beta^2+n^{-1}\int_0^t g^\tr\,\d Q\,g$,
in its turn leading to the natural and nearly unbiased estimator
$$\Bigl(\int_0^t g^\tr\,\d\hatt A_{II}\Bigr)^2
   -n^{-1}\int_0^t g(u)^\tr\,\d\hatt Q_n(u)\,g(u) $$
for $\beta^2$, where 
$$\d\hatt Q_n(u)=\pi_{II}
   \{G_n(u)^{-1}\,\d\hatt J_n(u)\,G_n(u)^{-1}\}\pi_{II}^\tr $$
is the empirical counterpart to $\d Q(u)$. 

These considerations lead to the risk estimator
$$\hatt R(I,t)=\hatt\mse_n(I,t)
   =\max\{\hatt\sqb(I,t),0\}
   +x_I^\tr\int_0^t G_{n,00}^{-1}\,\d\hatt J_{n,00}\,G_{n,00}^{-1}\,x_I, $$
where 
$$\hatt\sqb(I,t)=n\Bigl(\int_0^t b_{I,n}^\tr\,\d\hatt A_{II}\Bigr)^2
   -\int_0^t b_{I,n}^\tr\,\d\hatt Q_n\,b_{I,n}. $$

\vspace*{30pt}
\hrule
\section{The FIC and the Weighted FIC}
\label{section:fic}

Here we show how risk estimation methods developed above
lead to natural information criteria for model selection. 

The first such is a {\it focussed information criterion}
(FIC) that works for a given individual and a given
time point at which we wish optimal precision for 
his survival probability estimate. For the given 
covariate $x$ and time point $t$ we calculate 
\beq
\label{eq:fic}
\fic=\fic(I,x,t)=\max\{\hatt\sqb(I,x,t),0\}
   +\hatt\var(I,x,t) 
\eeq
for each candidate model $I$, where 
\beqn
\hatt\sqb(I,x,t)
&=&n\Bigl(\int_0^t b_{I,n}^\tr\,\d\hatt A_{II}\Bigr)^2
   -\int_0^t b_{I,n}^\tr\,\d\hatt Q_n\,b_{I,n}, \\
\hatt\var(I,x,t)
&=&x_I^\tr\int_0^t G_{n,00}^{-1}\,\d\hatt J_{n,00}\,
   G_{n,00}^{-1}\,x_I. 
\eeqn  
We note that $b_{I,n}(u)$ of (\ref{eq:bnu}) 
depends on $x$ and that the submatrices $G_{n,00}$ 
and so on of (\ref{eq:GntoG}) depend on $I$. 
In the end one selects the model 
with smallest value of the $\fic$ score number.

Note that FIC is sample-size dependent. In a situation
with a given amount of non-zero bias 
$\int_0^t \bar b_I^\tr\,\d A_{II}$, the $\hatt\sqb$
component of FIC will essentially increase with $n$,
whereas the variance component remains essentially 
constant. This goes to show that the best models
will tolerate less and less bias as $n$ increases,
and for sufficiently large $n$ only the full model
(which has zero modelling bias) will survive
FIC scrutiny. 

There are various variations on the FIC above. 
For a given individual who has survived 
up to time $t_1$ it is the conditional 
survival probabilities 
$$\Pr\{T^0\ge t_2\midd T^0\ge t_1,x\}
   =\exp\bigl[-\{H(t_2\midd x)-H(t_1\midd x)\}\bigr] $$
that are of interest. The development and formulae
above can be repeated mutatis mutandis with 
a given interval $[t_1,t_2]$ replacing $[0,t]$.
This gives a machinery for selecting models
that yield optimal estimation precision for 
conditional survival probabilities. It will also
be useful in many applications to monitor 
FIC scores for important candidate models
in terms of a `gliding time window', say $[t-\delta,t+\delta]$; 
successful models should then have good FIC scores
across time. We stress that it is not a paradox
that one model might be particularly good
at explaining the survival mechanisms involved
for short life-lengths, while another model 
might be much better for understanding 
the survival of the longer life-lengths. 
Our FIC takes this on board, and makes an explicit
model recommendation for each given time interval of interest.

Suppose now that a model is called for that works
well in an average sense across a given set of 
$(x,t)$ values, as opposed to a given $(x,t)$. 
Consider in general terms 
$$\EE_n(I)=n\int \{\tilda H_I(t\midd x)-H(t\midd x)\}^2
   \,\d w(t,x), $$
where $w(t,x)$ is a weight measure in the $(x,t)$ space.
This could for example take the form
\beq
\label{eq:EEwithK} 
\EE_n(I)=(1/K)\sum_{j=1}^K
   n\{\tilda H_I(t\midd x_j)-H(t\midd x_j)\}^2, 
\eeq
averaging across given covariate vectors 
$x_1,\ldots,x_K$. From (\ref{eq:basic}),
the random loss incurred using $I$ is
$$\EE_n(I)=\int\Bigl\{\rootn\int_0^t b_{I,n}(u,x)^\tr\,\d A_{II}(u)
   +x_I^\tr\int_0^t G_{n,00}(u)^{-1}\,\d V_{n,I}(u)\Bigr\}^2\,
   \d w(t,x), $$
writing now 
$$b_{I,n}(u,x)=G_{n,10}(u)G_{n,00}(u)^{-1}x_I-x_{II} $$
with explicit mention of $x$ in the notation. 

Its associated risk, the expected loss, is by previous
efforts closely approximated by 
the $w$-weighted risk 
$$R_n(I)=\E\int\Bigl[n\Bigl\{\int_0^t b_{I,n}(u,x)^\tr\,\d A_{II}(u)\Bigr\}^2
   +x_I^\tr\int_0^t G_{n,00}^{-1}\,\d J_{n,00}\,G_{n,00}^{-1}\,x_I\Bigr] 
   \,\d w(t,x). $$
We estimate the $w$-weighted squared bias and 
$w$-weighted variance contributions in turn. 
Define
\beqn
\whatsqb(I)
&=&n\int\Bigl\{\int_0^t b_{I,n}(u,x)^\tr\,\d\hatt A_{II}(u)\Bigr\}^2
   \,\d w(t,x) \\
& &\qquad\qquad
   -\,\int\int_0^tb_{I,n}(u,x)^\tr\,\d\hatt Q_n(u)\,
  b_{I,n}(u,x)\,\d w(t,x), 
\eeqn
which is an approximately unbiased estimator 
of the $w$-weighted squared bias term; and 
$$\whatvar(I)=\int\hatt\var(I,x,t)\,\d w(t,x). $$
Our $\wfic$ score, to be computed for each candidate model, is  
\beq
\label{eq:wfic}
\wfic(I)=\max\{\whatsqb(I),0\}+\whatvar(I).
\eeq
Again, in the end the model achieving the lowest $\wfic$ score
is selected. This scheme in particular gives rise 
to an algorithm associated with the (\ref{eq:EEwithK})
loss, weighting evenly across a finite set of covariate
vectors. 

A special case worth recording is when $t$ is fixed
and $w$ describes the covariate distribution. 
It is unknown, but may be approximated with the 
empirical distribution of covariates $x_1,\ldots,x_n$. 
This leads to $\wfic(I)$ as in (\ref{eq:wfic}) with 
\beqn
\whatvar(I)
&=&n^{-1}\sumin\hatt\var(I,x_i,t) \\
&=&\Tr\Bigl\{
   \Bigl(\int_0^t G_{n,00}^{-1}\,\d\hatt J_{n,00}\,G_{n,00}^{-1}\Bigr)
   \Bigl(n^{-1}\sumin x_{i,I}x_{i,I}^\tr\Bigr)\Bigr\}
\eeqn
whereas $\whatsqb(I)$ may be written 
\beqn
\sumin\{x_{i,I}^\tr\hatt B_I(t)
   -x_{i,II}^\tr\hatt A_{II}(t)\}^2 
   -n^{-1}\sumin
   \int_0^t b_{I,n}(u,x_i)^\tr\,\d\hatt Q_n(u)\,b_{I,n}(u,x_i), 
\eeqn
where 
$$\hatt B_I(t)=\int_0^t G_{n,00}(u)^{-1}G_{n,01}(u)\,\d\hatt A_{II}(u). $$

\medskip
{\sc Remark.}
Note that the $\wfic$ method as defined here is subtly
but crucially different from simply $w$-weighting of
the individual pointwise FIC scores, regarding 
how the truncation of the squared bias estimate 
is carried out. In (\ref{eq:wfic}), the truncation
to achieve nonnegativity of the estimate takes
place after the $w$-weighting, making it different
from $w$-weighting the collection of truncated 
$\sqb(I,x,t)$ terms. See in this connection also
Claeskens and Hjort (2007).~\blackbox

\vspace*{30pt}
\hrule
\section{Exact Risk Calculations}
\label{section:extra}

In the previous sections we were able to 
(i) develop formulae for risk functions 
and (ii) construct estimators for these. 
This led to model selection methods that may be
used in any given application. The present section
has a different aim, namely that of providing 
classes of case studies where the risk function
formulae can be computed explicitly, thereby
establishing a fair testing ground for 
model selection and model averaging methods. 
For reasons of space we shall be content to 
derive certain formulae under certain conditions, 
for biases and variances; these may then be
used to form concrete illustrations and test cases 
that for reasons of space can not be reported on 
in the present article. 

Assume that the components $x_1,\ldots,x_r$
of the covariate vector $x$ are distributed independently
of each other, with Laplace transforms 
$\E_*\exp(-\theta_j x_j)=\exp\{-M_j(\theta_j)\}$, say. 
Then 
$$\E_*\exp(-\theta^\tr x)
   =\exp\{-M_1(\theta_1)-\cdots-M_r(\theta_r)\}, $$
from which follows, taking second order derivatives
with respect to the $\theta$ components, that 
$$\E_*\exp(-\theta^\tr x)x_jx_k
   =\exp\Bigl\{-\sum_{l=1}^r M_l(\theta_l)\Bigr\}
   \{-M_j''(\theta_j)\delta_{j,k}+M_j'(\theta_j)M_k'(\theta_k)\}, $$
in terms of first and second order derivatives
of the $M_j$ functions. This implies that 
that the $r\times r$ limit function $G$ of (\ref{eq:GntoG}) 
may be expressed as 
$$G(u)=f(u)\{D(u)+z(u)z(u)^\tr\}C(u). $$
Here $f(u)=\exp\{-\sum_{l=1}^r M_l(A_l(u))\}$; 
$D(u)$ is the diagonal matrix with elements
$D_j(u)=-M_j''(A_j(u))$; and $z(u)$ is the vector
with elements $z_j(u)=M_j'(A_j(u))$. 
For a candidate set $I$ of covariates to include, 
the blocks of $G(u)$ can be read off from 
$$G(u)=f(u)C(u)\Bigl\{\pmatrix{D_0 &0 \cr 0 &D_1 \cr}
  +\pmatrix{z_0 \cr z_1 \cr}\pmatrix{z_0 \cr z_1 \cr}^\tr\Bigr\}, $$   
where $D_0$ and $D_1$ have components $D_j(u)$ 
where respectively $j\in I$ and $j\notin I$, 
and similarly $z_0$ and $z_1$ have components
$z_j(u)$ where $j\in I$ and $j\notin I$. In particular,
$$G_{00}(u)=f(u)C(u)(D_0+z_0z_0^\tr)
  \quadandquad
  G_{01}(u)=f(u)C(u)z_0z_1^\tr, $$
leading in turn, via the matrix inversion formula 
$$(D_0+z_0z_0^\tr)^{-1}
   =D_0^{-1}-{1\over 1+z_0^\tr D_0^{-1}z_0}
   D_0^{-1}z_0z_0^\tr D_0^{-1}, $$
to a formula for $G_{00}(u)^{-1}G_{01}(u)$ and then to
\beqn
b_I(u)
   &=&G_{10}(u)G_{00}(u)^{-1}x_I-x_{II} \\
&=&z_1z_0^\tr\Bigl(D_0^{-1}-{1\over 1+z_0^\tr D_0^{-1}z_0}
   D_0^{-1}z_0z_0^\tr D_0^{-1}\Bigr)x_I-x_{II} \\
&=&z_1{z_0^\tr D_0^{-1}x_I\over 1+z_0^\tr D_0^{-1}z_0}-x_{II}. 
\eeqn 

Assume for a concrete example that $x_j\sim\Gam(a_j,b_j)$
for $j=1,\ldots,r$, for which the Laplace transforms
are $\{b_j/(b_j+\theta_j)\}^{a_j}$ with 
$M_j(\theta_j)=a_j\log(1+\theta_j/b_j)$. Then
$$M_j'(\theta_j)={\xi_j\over 1+\theta_j/b_j}
  \quadandquad
  M_j''(\theta_j)=-{\xi_j/b_j\over (1+\theta_j/b_j)^2}, $$
with $\xi_j=\E_*x_j=a_j/b_j$. This yields a bias
function $b_I(u)$ with components 
$$b_{I,j}(u)={g_I(u)\over 1+\sum_{j\in I}b_j\xi_j}
   {\xi_j\over 1+A_j(u)/b_j}-x_j 
  \quad {\rm for\ }j\in II=I^c, $$
where $g_I(u)=\sum_{j\in I}\{b_j+A_j(u)\}x_j$. It follows
that the important bias component of (\ref{eq:basic}) 
may be written 
$$\rootn\int_0^t b_I^\tr\,\d A_{II}
  =\rootn\Bigl\{\int_0^t{g_I(u)\over 1+\sum_{j\in I}b_j\xi_j}
   \sum_{j\in II}{\xi_j\alpha_j(u)\over 1+A_j(u)/b_j}\,\d u
  -x_{II}^\tr A_{II}(t)\Bigr\}. $$
These bias functions are easily computed and displayed,
for given covariate distributions and given
hazard regression functions. 

To handle the variance part of (\ref{eq:sqbplusvar})
we need an explicit formula for $\d J(u)$ and then for 
$$G_{00}^{-1}\,\d J_{00}\,G_{00}^{-1}
  \quadandquad
  G(u)^{-1}\,\d J(u)\,G(u)^{-1}. $$
We start with
$$\E_*\exp(-s\theta^\tr x)x_jx_k
   =\exp\Bigl\{-\sum_{l=1}^rM_l(s\theta_l)\Bigr\}
   \{-M_j''(s\theta_j)\delta_{j,k}
   +M_j'(s\theta_j)M_k'(s\theta_k)\}, $$
and then take the derivative w.r.t.~$s$, 
and set $s=1$ in the resulting equations. This yields
\beqn
\E_*\exp\{-\theta^\tr x)x_jx_k\,\theta^\tr x
&=&f^*(\theta)\bigl[
  \{M_j'''(\theta_j)\theta_j-g^*(\theta)M_j''(\theta_j)\}\delta_{j,k} \\
& &\qquad
-\,M_j'(\theta_j)M_k''(\theta_k)\theta_k
-M_j''(\theta_j)M_k'(\theta_k)\theta_j \\
& &\qquad
+\,g^*(\theta)M_j'(\theta_j)M_k'(\theta_k)\bigr], 
\eeqn
where
\beqn
f^*(\theta)
   =\exp\Bigl\{-\sum_{l=1}^r M_l(\theta_l)\Bigr\}
\quadandquad
g^*(\theta)
   =\sum_{l=1}^r M_l'(\theta_l)\theta_l. 
\eeqn

Let now $A_j(t)=\alpha_jt$ for $j=1,\ldots,r$,
i.e.~the $\alpha_j$ regressor functions are taken constant. 
The above leads with some further work to a formula for 
$$\E_*\exp\{-x^\tr A(u)\}xx^\tr\,x^\tr\,\d A(u)
   =f(u)\{E(u)+F(u)\}\,\d u $$
where the $E(u)$ and $F(u)$ matrix functions 
are described below; 
also, $f(u)=\exp\{-\sum_{l=1}^rM_l(A_l(u))\}$
is as for the bias calculations above. 
The $E(u)$ is diagonal with elements 
$$E_j(u)=M_j'''(A_j(u))\alpha_j-g(u)M_j''(A_j(u)), $$ 
where $g(u)=\sum_{l=1}^r M_l'(A_l(u))\alpha_l$. 
Next, $F(u)$ has $(j,k)$ element 
\beqn
-M_j'(A_j(u))M_k''(A_k(u))\alpha_k
  &-&M_j''(A_j(u))M_k'(A_k(u))\alpha_j \\
& &\qquad  +\,g(u)M_j'(A_j(u))M_k'(A_k(u)).
\eeqn
These results may be used to compute the variance terms
$$x_I^\tr\int_0^t G_{00}^{-1}\,\d J_{00}\,G_{00}^{-1}\,x_I $$
and thereby the mean squared errors for different 
candidate models. 
These formulae may in particular be used for the 
case mentioned earlier, with independent
$\Gam(a_j,b_j)$ distribution for the $x_j$ components,
and for which 
$M_j'''(\theta_j)=2(\xi_j/b_j^2)/(1+\theta_j/b_j)^3$. 

Various concrete illustrations may now be given,
for the specific case of independent gamma distributed
covariates, to exhibit and examine various aspects 
and issues involved in model selection and model averaging. 
These relate in various ways to modelling bias
versus estimation variance. We may for example 
show that when $\alpha_j(u)$s are small in size,
then it may be best not to include these in the
selected model, depending also on the sizes 
of $x_I$ and $x_{II}$. We would also be able
to illustrate how the complexity of the best model
increases with higher sample size, and how 
the qualitative results depend on the relative spread 
of the distributions of covariates. 

\vspace*{30pt}
\hrule
\section{Concluding Remarks}
\label{section:concludingremarks}

Here we offer some concluding comments, some pointing
to natural extensions of the material and methods
we have presented above. 

\hop
(1) In a planned extended version of this paper
space will be given to analysis of a real data set
and to instructive simulation setups. 

\hop
(2) We have throughout used `vanilla weights' 
for the Aalen estimators $\hatt A$ of (\ref{eq:aalen}).
With more sophisticated weighting the estimator
$$\hatt A(t,k)=\int_0^t
   \Bigl\{n^{-1}\sumin Y_i(u)k_i(u)x_ix_i^\tr\Bigr\}^{-1}
   n^{-1}\sumin x_ik_i(u)\,\d N_i(u) $$
may perform slightly better; see 
Huffer and McKeague (1991). Also for such schemes
a FIC and wFIC methodology may be developed,
generalising methods given in the present paper. 

\hop
(3) A local asymptotic framework may be put up
for the Aalen model, similar in spirit 
to that employed in Hjort and Claeskens (2003)
and Claeskens and Hjort (2003) for purely 
parametric models. Here one would use 
hazard rates 
$$h_i(u)=\sum_{j=1}^p x_{i,j}\alpha_j(u)
   +\sum_{j=1}^q z_{i,j}\delta_j(u)/\rootn, $$
with $x_{i,j}$s protected covariates considered
important to include in all candidate models,
and $z_{i,j}$s the potentially discardable ones.
A precise asymptotic description may now be given
of all limiting risk functions, in terms of 
the $\delta_1,\ldots,\delta_q$ functions. 

\hop
(4) A fair question to ask is the behaviour
of the final estimator, say 
$$H^*(t\midd x)=\tilda H_{\hatt I}(t\midd x), $$
where $\hatt I$ is the data-dependent set
of finally included covariates. This is 
a complicated question without any easy answer.
Inside the local asymptotic framework of (3) 
methods of Hjort and Claeskens (2003) may 
be used to describe the limit distribution 
of $\rootn\{H^*(t\midd x)-H(t\midd x)\}$,
in terms of a non-linear mixture of biased normals. 
This also opens the door to general model average
strategies, as opposed to limiting inference
methods to those that rely on deciding on 
only one model. 

\hop
(5) We have developed machinery for answering
the question `should covariate $j$ be included
in the nonparametric Aalen model, or not?'. 
More ambitiously and more laboriously, one
can give not only two but three potential
outcomes for each covariate: 
it might be excluded;
it might be included nonparametrically;
it might be included parametrically. 
The latter possibility refers e.g.~to the 
model where $\alpha_j(u)$ is constant; 
see McKeague and Sasieni (1994) for 
treatment of such models. Again a FIC and 
a wFIC apparatus may be developed, 
requiring however more mathematical vigour. 

\hop 
{\bf Acknowledgements.} 
This paper reports on work carried out in the stimulating
environment of the Centre of Advanced Study 
at the Norwegian Academy of Science and Letters, 
where \O rnulf Borgan and Odd Aalen organised 
a research group on Statistical Analysis of Complex 
Event History Data. Conversations with Axel Gandy
were particularly fruitful. 

\vspace*{30pt}
\hrule

\def\jasa{Journal of the American Statistical Association}
\def\et{Econometric Theory}
\def\biometrika{Biometrika}

\section*{References}


\begin{enumerate}

\item Andersen, P.K., Borgan, \O., Gill, R.~and Keiding, N. (1993).
{\it Statistical Models Based on Counting Processes.}
Springer-Verlag, Heidelberg. 

\item Claeskens, G.~and Hjort, N.L. (2003).
The focused information criterion
[with discussion], 
{\it\jasa}, {\bf 98}, 900--916 and 938--945.

\item Claeskens, G.~and Hjort, N.L. (2007).
Minimising average risk in regression models.
{\it\et}, to appear. 

\item Hjort, N.L.~and Claeskens, G. (2003).
Frequentist average estimators [with discussion], 
{\it\jasa}, {\bf 98}, 879--899 and 938--945.

\item Hjort, N.L.~and Claeskens, G. (2006).
Focused information criteria and model averaging
for the Cox hazard regression model,
{\it\jasa}, {\bf 101}, 1449--1464.

\item Huffer, F.~and McKeague, I.W. (1991).
Weighted least squares estimation for Aalen's additive
risk model.
{\it\jasa}, {\bf 86}, 114--129. 

\item McKeague, I.W.~and Sasieni, P.D. (1994).
A partly parametric additive risk model,
{\it\biometrika}, {\bf 81}, 501--514. 

\item Martinussen, T.~and Scheike, Thomas H. (2006). 
{\it Dynamic Regression Models for Survival Data.}
Springer-Verlag, New York. 

\item Aalen, O.O. (1980).
A model for nonparametric regression analysis
of counting processes, 
in {\it Mathematical Statistics and Probability Theory}
(eds.~W.~Klonecki, A.~Kozek and J.~Rosinski).  
Proceedings of the 6th international conference,
Wisla, Poland, 1--25. 

\end{enumerate}

\end{document}